\documentclass[journal=jacsat,manuscript=article]{achemso}
\usepackage{chemformula} 
\usepackage[T1]{fontenc} 

\usepackage{subcaption,tikz,float,tabularx}
\usepackage{siunitx}
\usetikzlibrary{spy,decorations.fractals}

\graphicspath{{./figure-exp3/}{./figure-exp1/}{./figure-exp2/}{./figure-exp4/}{./figure-author/}}

\definecolor{spy_color_red}{RGB}{49,120,136}    
\definecolor{spy_color_orange}{RGB}{255, 255, 255}   
\definecolor{mark_color}{RGB}{255,255,255}

\author{Jie Zhou}
\affiliation[Sichuan University]
{College of Electronics and Information Engineering, Sichuan University, Chengdu 610065, China}

\author{Shuyang Xie}
\affiliation[The Hong Kong University of Science and Technology]
{Thrust of Microelectronics of Function Hub, The Hong Kong University of Science and Technology
(Guangzhou), Guangzhou 511400, China}

\author{Yang Wu}
\affiliation[Sichuan University]
{College of Electronics and Information Engineering, Sichuan University, Chengdu 610065, China}

\author{Lei Jiang}
\affiliation[Sichuan University]
{College of Electronics and Information Engineering, Sichuan University, Chengdu 610065, China}

\author{Yimou Luo}
\affiliation[Sichuan University]
{College of Electronics and Information Engineering, Sichuan University, Chengdu 610065, China}

\author{Jun Wang}
\email{jwang@scu.edu.cn}
\affiliation[Sichuan University]
{College of Electronics and Information Engineering, Sichuan University, Chengdu 610065, China}

\title[An \textsf{achemso} demo] {Eyepiece-free pupil-optimized holographic near-eye displays}


\begin{document}

\begin{abstract}
Computer-generated holography (CGH) represents a transformative visualization approach for next-generation immersive virtual and augmented reality (VR/AR) displays, enabling precise wavefront modulation and naturally providing comprehensive physiological depth cues without the need for bulky optical assemblies. Despite significant advancements in computational algorithms enhancing image quality and achieving real-time generation, practical implementations of holographic near-eye displays (NEDs) continue to face substantial challenges arising from finite and dynamically varying pupil apertures, which degrade image quality and compromise user experience. In this study, we introduce an eyepiece-free pupil-optimized holographic NED. Our proposed method employs a customized spherical phase modulation strategy to generate multiple viewpoints within the pupil, entirely eliminating the dependence on conventional optical eyepieces. Through the joint optimization of amplitude and phase distributions across these viewpoints, the method markedly mitigates image degradation due to finite pupil sampling and resolves inapparent depth cues induced by the spherical phase. The demonstrated method signifies a substantial advancement toward the realization of compact, lightweight, and flexible holographic NED systems, fulfilling stringent requirements for future VR/AR display technologies.
\end{abstract}

\section{Introduction}\label{sec:Introduction}
Computer-generated holography (CGH) has emerged as a transformative visual technology poised to drive next-generation immersive virtual and augmented reality (VR/AR) experiences through precise reconstruction of complete optical wavefronts \cite{dong2025motion, Li2025LightSciAppl,sui2024non, yu2023ultrahigh, pi2022review}.  Unlike traditional stereoscopic displays, which approximate depth cues through parallax, holographic displays inherently reproduce comprehensive physiological depth information, including continuous accommodation and realistic focus effects, without requiring bulky optical components \cite{fu2024photonics,yu2022dynamic,wang2022high,feng2019spin}. This direct wavefront-level modulation bestows holographic near-eye displays (NEDs) with unprecedented capabilities, such as ultra-high resolution, accurate reproduction of focus cues, and customizable aberration correction, within a significantly compact form factor—features ideal for advanced VR/AR applications \cite{xia2025off,tseng2024neural, gopakumar2024full, chen2024ultrahigh, kim2024holographic, tian2024srgan}. Recent, breakthroughs of powerful computational methods have driven significant advancements in CGH algorithms, delivering substantial improvements in holographic image quality \cite{zhang20173d, peng2020neural, xia2023investigating, choi2021neural, chakravarthula2020learned} and real-time computational efficiency \cite{shi2021towards,liu20234k,zhu2023computer,dong2023fourier,yu2023asymmetrical,zhong2023real}, overcoming previous bottlenecks that hindered practical deployment. Despite these compelling advantages, holographic NEDs continue to face substantial practical challenges associated with finite and dynamically varying pupil apertures of the human eye. The eye's pupil functions as an adaptive aperture that partially samples the reconstructed holographic wavefront. Thus, any misalignment or incomplete sampling directly leads to a noticeable degradation of image quality and perceptual realism. Addressing this pupil-sampling challenge remains an essential prerequisite for realizing practical holographic NED systems.

Recent advances in CGH have begun addressing pupil-dependent constraints by integrating viewer-specific parameters into hologram generation frameworks. Chakravarthula et al. \cite{chakravarthula2022pupil} introduced pupil-aware holography, a computational framework optimizing hologram according to pupil aperture and position, ensuring robust visual performance across the entire eyebox region. Building upon this concept, Shi et al. \cite{shi2024ergonomic} proposed ergonomic-centric holography, enhancing depth realism through realistic defocus effects, supporting broad eyebox adaptation, and enabling filter-free optical designs by effectively managing high-order diffractions. Similarly, Wang et al. \cite{wang2024pupil} investigated dynamic pupil shapes and motions, proposing adaptive depth-of-field holography to closely replicate natural incoherent depth perception within coherent holographic display systems. Relay-optics-based solutions have also been explored to address pupil sampling limitations. Methods such as multi-viewpoint optimization utilizing pupil masks \cite{chen2024multiple1} and multi-sub-hologram optimization based on viewing windows \cite{chen2024multiple2} enhance image consistency and quality across various pupil positions. However, these methods typically depend on complex optical elements, including eyepieces or relay lenses, thereby increasing system complexity, form-factor bulkiness, and susceptibility to optical aberrations. This poses significant challenges to the practical deployment of lightweight, wearable NED systems.

In this study, we present an eyepiece-free pupil-optimized holographic NED that overcomes these critical limitations by relying solely on wavefront modulation tailored to the user’s pupil. Our method uniquely leverages customized spherical phase modulations to generate multiple viewpoints within the pupil, entirely eliminating the reliance on conventional eyepiece optics. Jointly optimizing amplitude and phase distributions across these viewpoints significantly reduces image degradation due to finite pupil sampling while effectively resolving the inapparent depth cues induced by spherical phases. 
Moreover, we present an advanced optimization-based phase-only hologram (POH) encoding technique that efficiently converts complex amplitude hologram (CAH) into a high-quality, artifact-suppressed phase-only representation suitable for practical optical implementation of eyepiece-free holographic NED. Extensive numerical simulation and optical experiments validate our design, consistently delivering high-quality, stable 3D imagery over the full range of pupil apertures. This work represents a significant step toward the development of compact, lightweight, and flexible holographic NEDs that meet the stringent requirements of next-generation VR/AR applications.

\section{Problem analysis}\label{sec:problem analysis}
\begin{figure*}[h!]
    \centering
    \includegraphics[width=6.4in]{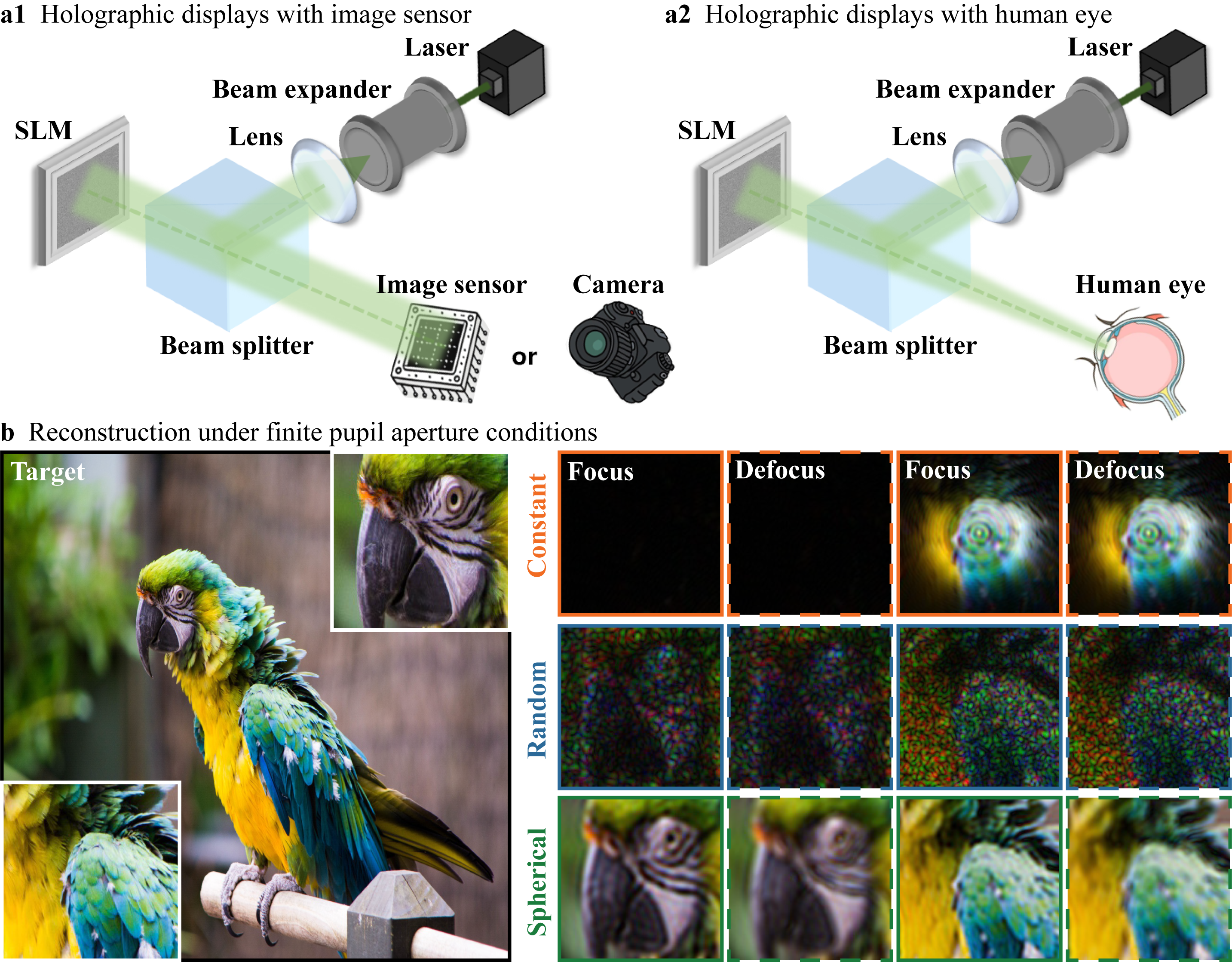}
    \caption{Experimental setups and impact of finite pupil aperture on holographic reconstruction. (a) Schematic illustration of holographic display configurations: (a1) Conventional holographic capture employing an image sensor or camera; (a2) Direct human-eye holographic viewing setup. (b) Comparison of holographic reconstruction under finite pupil aperture conditions for holograms with constant, random, and spherical phases.}
    \label{fig1}
\end{figure*}

Figure \ref{fig1}(a1) illustrates a conventional experimental configuration for capturing reconstructed holographic images. In this setup, an image sensor is positioned along the optical axis to record intensity distributions at a specified plane, which can be shifted axially to acquire intensity patterns at multiple depths. Typically, large-aperture lenses such as digital single lens reflex (DSLR) camera lenses are employed to collect nearly the complete emitted wavefront from holographic displays, ensuring high-quality image reconstruction on the sensor. However, this routine practice does not adequately replicate the viewing conditions experienced by human observers, whose visual perception is mediated through an eye with a finite pupil aperture, significantly constraining the wavefront portion reaching the retina (Figure \ref{fig1}(a2)).

Figure \ref{fig1}(b) demonstrates that holograms with uniform phase distributions suffer substantial degradation in perceived image completeness under finite pupil aperture conditions due to inherently restricted angular divergence. Wavefronts dominated by low spatial frequency components propagate within narrow angular distributions, thus delivering incomplete visual information to the retina through a finite pupil aperture. Increasing pupil aperture theoretically enhances information throughput, yet physiological constraints such as pupil constriction under bright illumination inherently limit this method's practical utility. 
Introducing random phase distributions mitigates these angular constraints by broadening the wavefront divergence angle, thus improving information throughput. However, random phases significantly exacerbate speckle noise, deteriorating image quality, especially under conditions of reduced pupil aperture. 
Conventional optical solutions such as the incorporation of refractive lenses can focus wavefronts onto the pupil but inherently introduce aberrations and system complexity, limiting practical applicability in compact NEDs.
Spherical phase holograms present an alternative strategy, enhancing image completeness by effectively converging emitted wavefronts onto the pupil and reducing speckle noise. Nevertheless, this method exhibits two critical limitations. 
Firstly, because of the convergence of the spherical wave at the pupil plane, this plane effectively serves as the Fourier domain of the reconstructed image. Consequently, the pupil aperture acts as a spatial frequency filter, preferentially admitting low spatial frequency components while attenuating or excluding higher spatial frequency details, thus diminishing overall visual fidelity. 
Secondly, spherical phase holograms inherently compromise essential depth perception cues. Under typical human viewing conditions, defocus blur provides a crucial indicator of relative object distances, significantly contributing to depth perception. However, spherical phase holograms maintain uniform sharpness across extended axial ranges, thereby suppressing defocus-related depth cues and negatively impacting the user's perception of authentic 3D structures.
\section{Method}\label{sec:Method}
\begin{figure*}[h!]
    \centering
    \includegraphics[width=6.4in]{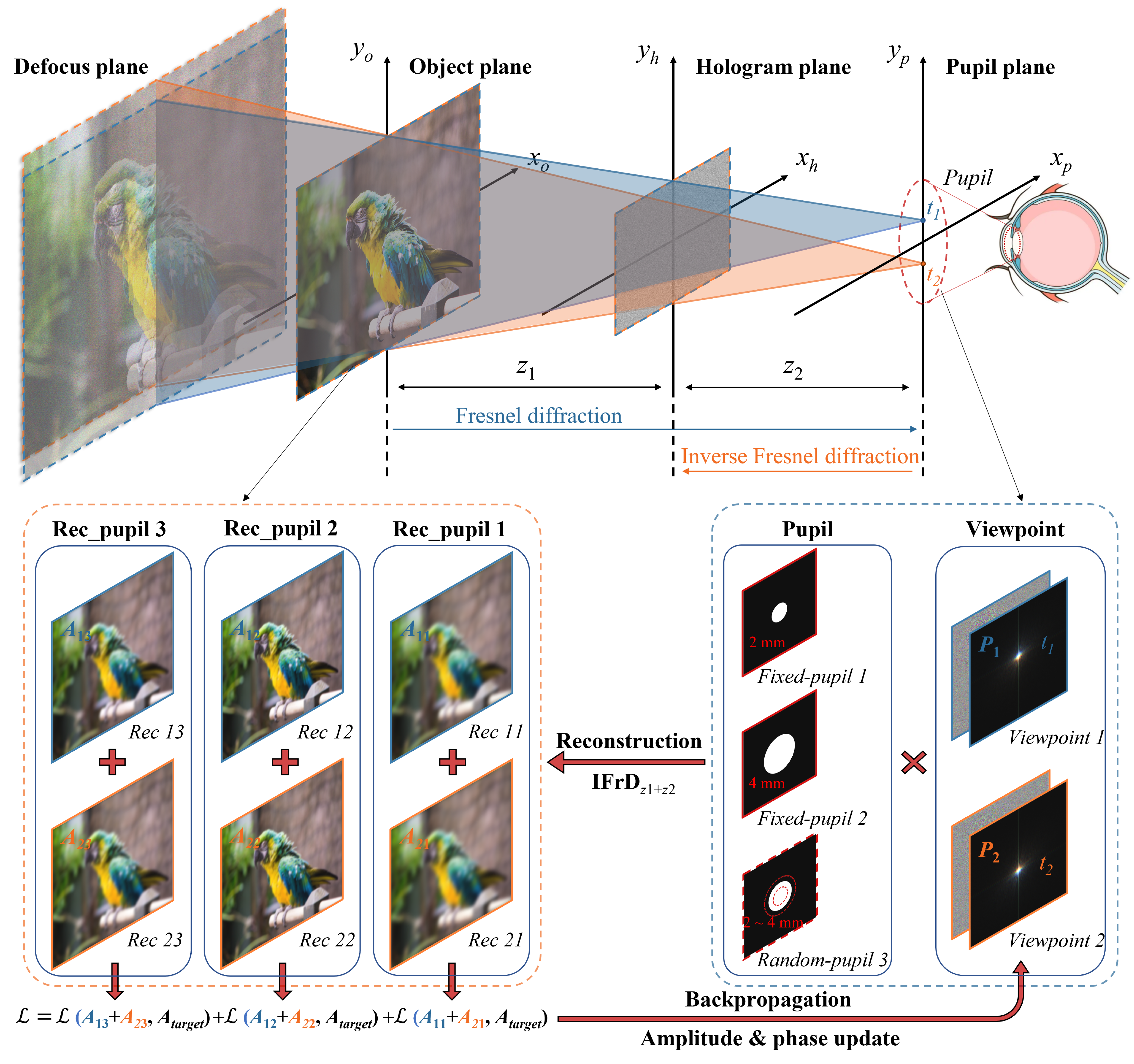}
    \caption{Schematic illustration of eyepiece-free pupil-optimized holographic NED. The display architecture employs a customized spherical phase modulation strategy to generate multiple viewpoints within the pupil, thereby eliminating the need for conventional eyepiece optics. By jointly optimizing the amplitude and phase distributions across these viewpoints, the method significantly mitigates quality degradation due to finite pupil sampling and resolves the inapparent depth cues induced by the spherical phase.}
    \label{fig2}
\end{figure*}

To address these inherent limitations, we propose an eyepiece-free pupil-optimized holographic NED, as illustrated in Figure \ref{fig2}. The novel method introduces multiple viewpoints within the finite pupil aperture and employs joint optimization of amplitude and phase distributions across these viewpoints, substantially enhancing both reconstruction quality and depth perception.
In this configuration, rays originating from identical object points simultaneously enter the pupil through distinct viewpoints. Consequently, rays fail to converge accurately on the retina when the user's eye is unfocused, reinforcing depth perception cues, whereas precise convergence on the retina occurs upon correct focusing, with complementary information from each viewpoint further enhancing image quality.

To generate multiple viewpoints within the pupil, the target 3D image $A_t(x_o,y_o)$ is initially modulated by several convergent spherical phases. This modulation produces a complex amplitude distribution given by:
\begin{equation}
O_i({x_o},{y_o}) = A_t({x_o},{y_o}) \cdot \exp \left\{ \frac{{ - jk[(x_o-x_i)^2 + (y_o-y_i)^2]}}{{2({z_1} + {z_2})}} \right\},
\end{equation}
where $(x_i, y_i)$ denotes the position of the 
$i$-th viewpoint, $k = 2\pi/\lambda$ is the wave number, $z_1$ is the distance from the object plane to the hologram plane, and $z_2$ is the distance from the hologram plane to the pupil plane. The complex amplitude $P_i(x_p, y_p)$ of the viewpoints on the pupil plane is obtained through the first-step Fresnel propagation:
\begin{equation}
P_i({x_p},{y_p}) = \exp \left[\frac{{jk(x_p^2 + y_p^2)}}{{2({z_1} + {z_2})}}\right]\mathcal{F}\left\{ O_i\cdot\exp\left[\frac{{jk(x_o^2 + y_o^2)}}{{2({z_1} + {z_2})}}\right]\right\},
\end{equation}
where $\mathcal{F}$ denotes the Fourier transform. The complex distribution $H_i(x_h, y_h)$ of the viewpoints on the hologram plane is then calculated through the second-step inverse Fresnel propagation:
\begin{equation}
H_i({x_h},{y_h}) = \exp \left[\frac{{jk(x_h^2 + y_h^2)}}{{-2{z_2}}}\right]\mathcal{F}^{-1}\left\{ P_i\cdot\exp\left[\frac{{jk(x_p^2 + y_p^2)}}{{-2 {z_2}}}\right]\right\}.
\end{equation}
The reconstructed amplitude ${A_{ik}}({x_o},{y_o})$ for viewpoints with different pupil apertures is obtained as follows:
\begin{equation}
A_{ik}(x_o, y_o) = \left| \exp \left[ \frac{jk(x_o^2 + y_o^2)}{-2(z_1 + z_2)} \right] \right. \left. \mathcal{F}^{-1} \left\{ P_i \cdot M_k \cdot \exp \left[ \frac{jk(x_p^2 + y_p^2)}{-2(z_1 + z_2)} \right] \right\} \right|,
\end{equation}
where ${M_k}$ represents the mask of the $k$-th pupil.

Due to coherence-induced interference among viewpoints, we utilize a time-division multiplexing (TDM) technique. By sequentially displaying holograms for each viewpoint faster than the persistence of human vision, interference effects are eliminated. The integrated perceived amplitude across viewpoints becomes:
\begin{equation}
A_{k}(x_o, y_o) = \sum\limits_i A_{ik}(x_o, y_o).
\end{equation}

To determine the optimal complex amplitude $P_i = A_{p_i}\exp(j\Phi_{p_i})$ of each viewpoint on the pupil plane, we introduce a multi-aperture pupil optimization algorithm. By jointly optimizing the amplitudes and phases across all viewpoints, the method aims to combine their contributions to achieve high-quality reconstruction across various pupil apertures. We adopt a hybrid strategy utilizing a combination of random and fixed aperture pupils, ensuring that high-quality images remain perceptible within the normal human pupil aperture range. The optimization process is mathematically formulated as the following minimization problem:
\begin{equation}
\mathop {{\rm{argmin}}}\limits_{A_{p_i}, \Phi_{p_i}, s_p} \sum\limits_k{{\cal L}}(s_p \cdot \sum\limits_i{{A}_{ki}},{A_t}),
\end{equation}
where $A_t$ represents the desired target amplitude, $s_p$ is the global scale factor, and ${{\cal L}}$ denotes the loss function guiding the optimization process. The loss function ${{\cal L}}$ is designed as a weighted sum of two distinct components that capture various quality aspects of the reconstructed image:
\begin{equation}
\mathcal{L} = \lambda_{\ell_2} \mathcal{L}_{\ell_2} + \lambda_{SSIM} \mathcal{L}_{SSIM},
\end{equation}
where $\mathcal{L}_{\ell_2}$ and $\mathcal{L}_{SSIM}$ represent the ${\ell_2}$-norm loss and the Structure Similarity Index Measure (SSIM) loss, respectively. 
$\mathcal{L}_{\ell_2}$ helps ensure that the overall amplitude distribution of the reconstructed image closely matches the desired target. It serves as a fundamental measure for the global reconstruction accuracy.
$\mathcal{L}_{SSIM}$ enhances the quality of the reconstructed image by preserving spatial relationships.
The parameters $\lambda_{\ell_2}$ and $\lambda_{SSIM}$ control the relative importance of each loss component in the optimization process.

After the optimal pupil-plane solution $P_i$ is obtained, it is numerically propagated to the hologram plane to generate a complex amplitude hologram (CAH). For practical holographic displays, the CAH must be encoded as either a phase-only hologram (POH) or an amplitude-only hologram (AOH). Traditional interference-based encoding for AOH often introduces significant direct current (DC) and conjugate noise, leading to low diffraction efficiency \cite{wang2022simultaneous, wang2023cross, yang2022diffraction, yang2024direct}. In contrast, POHs tend to offer higher diffraction efficiency, making them more suitable for high-quality holographic reconstructions.
Although the error diffusion method (EDM) \cite{tsang2013novel} and the double phase method (DPM) \cite{mendoza2014encoding} are well established for POH generation, they exhibit limitations for eyepiece-free holographic NED: EDM performance deteriorates in the presence of large phase variations, whereas DPM introduces optical artifacts that compromise the viewing experience.
To address these challenges, we propose an optimization-based phase-only encoding technique. The objective is to determine the optimized phase pattern $H_i=\exp(j\Phi_{h_i})$ of each viewpoint on the hologram plane, which can be reformulated as the following optimization problem:
\begin{equation}
\mathop {{\rm{argmin}}}\limits_{\Phi_{h_i}} {{\cal L}}_{\ell_2}({I}_{r_i}\cdot M_{k_{max}},{I_{p_i}}\cdot M_{k_{max}})+{{\cal L}}_{\ell_2}({R}_{r_i}\cdot M_{k_{max}},{R_{p_i}}\cdot M_{k_{max}}),
\end{equation}
where ${R}_{r_i}$, ${I}_{r_i}$ and ${R}_{p_i}$, ${I}_{p_i}$ are real and imaginary components of reconstructed and target complex amplitudes $P_i$ of each viewpoint on the pupil plane, respectively, $M_{k_{max}}$ is the max aperture pupil mask.
\section{Results}\label{sec:Experiment}
\begin{figure*}[p!]
    \centering
    \includegraphics[width=6.4in]{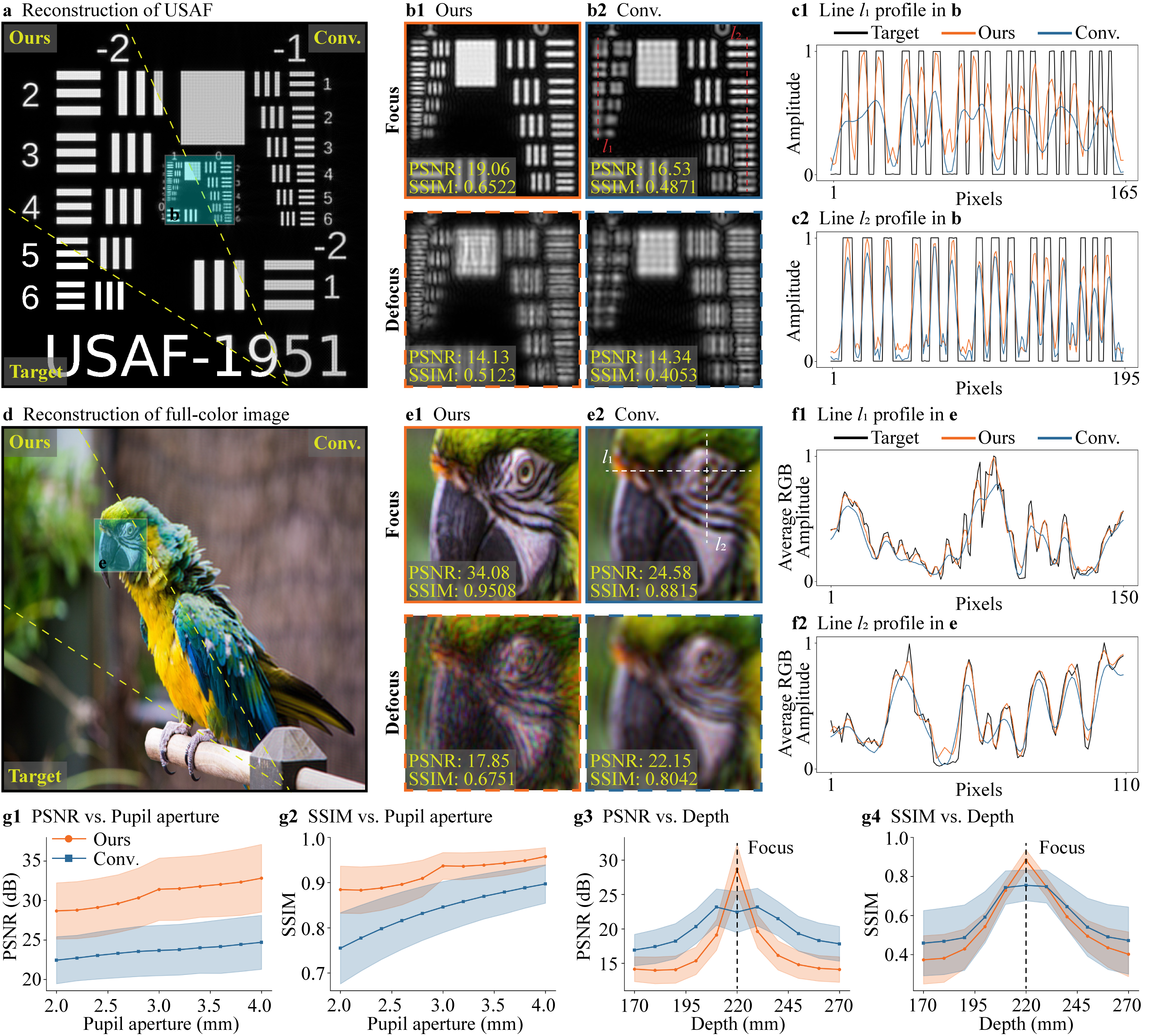}
    \caption{Comprehensive numerical evaluation and benchmarking of the proposed holographic NED method. 
    (a) Comparative amplitude reconstructions of the USAF-1951 resolution target using the proposed method ("Ours") and conventional spherical phase holography ("Conv.") under a pupil aperture of \SI{2}{\milli\meter}. 
    (b) Detailed visual comparison of reconstructed amplitude distributions at the focus plane (\SI{220}{\milli\meter}) and defocus plane (\SI{240}{\milli\meter}) corresponding to the region marked by the blue square in (a): (b1) Proposed method; (b2) Conventional method.
    (c) Quantitative amplitude profiles along representative red line sections indicated in (b): (c1) Profile of line $l_1$; (c2) Profile of line $l_2$. (d) Demonstration of full-color image reconstruction capability. 
    (e) Detailed visual comparison of the reconstructed full-color image at the focus plane and defocus plane corresponding to the region marked by the blue square in (d): (e1) Proposed method; (e2) Conventional method. 
    (f) Quantitative average RGB amplitude profiles along representative white line sections indicated in (e): (f1) Profile of line $l_1$; (f2) Profile of line $l_2$. 
    (g) Statistical evaluation of reconstruction quality metrics over 100 images from the DIV2K \cite{agustsson2017ntire} dataset: (g1,g2) Analysis across varying pupil apertures; (g3,g4) Analysis across different depths.}
    \label{fig3}
\end{figure*}

A comprehensive numerical evaluation of the proposed holographic NED method was systematically conducted and rigorously benchmarked against the conventional spherical phase holography method. As illustrated in Figure \ref{fig3}(a), employing a USAF-1951 resolution target, CAHs were generated at a spatial resolution of $1080 \times 1080$ pixels with a pixel pitch of \SI{8}{\micro\meter}, illuminated by the light source at \SI{671}{\nano\meter}. The visual results are evaluated under a pupil aperture of \SI{2}{\milli\meter}. Propagation distances were precisely set, with $z_1$ (object-to-hologram plane) and $z_2$ (hologram-to-pupil plane) configured at \SI{80}{\milli\meter} and \SI{140}{\milli\meter}, respectively, yielding an focus depth of \SI{220}{\milli\meter}. To evaluate defocus performance quantitatively, an additional evaluation plane was positioned at \SI{240}{\milli\meter}. Three distinct viewpoints were considered in the evaluation. Fixed aperture pupils of \SI{2}{\milli\meter}, \SI{3}{\milli\meter}, and \SI{4}{\milli\meter} were used, supplemented by two random aperture pupils. The parameters $\lambda_{\ell_2}$ and $\lambda_{SSIM}$ were set to 0.95 and 0.05, respectively. The optimization process employed a learning rate of 0.02 and was conducted over 2000 iterations to ensure convergence. Image quality metrics included Peak Signal-to-Noise Ratio (PSNR) and SSIM.

Detailed visual comparisons provided in Figure \ref{fig3}(b) demonstrate substantial enhancements in reconstructed image quality using the proposed method, particularly in preserving high-frequency spatial details compared to the conventional method. 
Critically, under defocus conditions, the proposed method generates more pronounced blur characteristics, effectively enhancing depth perception.
Figure \ref{fig3}(c) shows quantitative amplitude profiles extracted along the red line sections in Figure \ref{fig3}(b) that further confirm the superior capability of the proposed method in preserving fine structural details and edge fidelity compared to conventional methods. The enhanced similarity between the reconstructed and target amplitude profiles provides robust quantitative validation of the proposed method's effectiveness.

Expanding the evaluation to full-color holographic reconstructions at wavelengths of \SI{473}{\nano\meter}, \SI{532}{\nano\meter}, and \SI{671}{\nano\meter}, Figure \ref{fig3}(d) maintains identical experimental conditions as previously described.
Detailed visual comparisons (Figure \ref{fig3}(e)) indicate markedly improved visual clarity and color fidelity at the focus plane with the proposed method. Furthermore, this method distinctly produces a more significant blur away from the focus plane, enhancing depth contrast. Corresponding average RGB amplitude profile comparisons (Figure \ref{fig3}(f)) further affirm the method's robustness in accurately reconstructing intricate spatial features and high-frequency color details.

Statistical evaluations performed on a diverse set of 100 randomly selected images from the DIV2K \cite{agustsson2017ntire} dataset (Figure \ref{fig3}(g)) consistently demonstrate the superiority of the proposed method over the conventional holography method across realistic pupil aperture ranges. Specifically, statistically significant enhancements in PSNR and SSIM are evident (Figures \ref{fig3}(g1,g2)). Furthermore, depth-dependent performance analyses (Figures \ref{fig3}(g3,g4)) indicated a rapid decline in PSNR and SSIM outside the focus plane, substantiating the proposed method’s effectiveness in enhancing depth-dependent blur, a key factor for achieving authentic 3D visualization and improved depth realism.

\begin{figure*}[h!]
    \centering
    \includegraphics[width=6.4in]{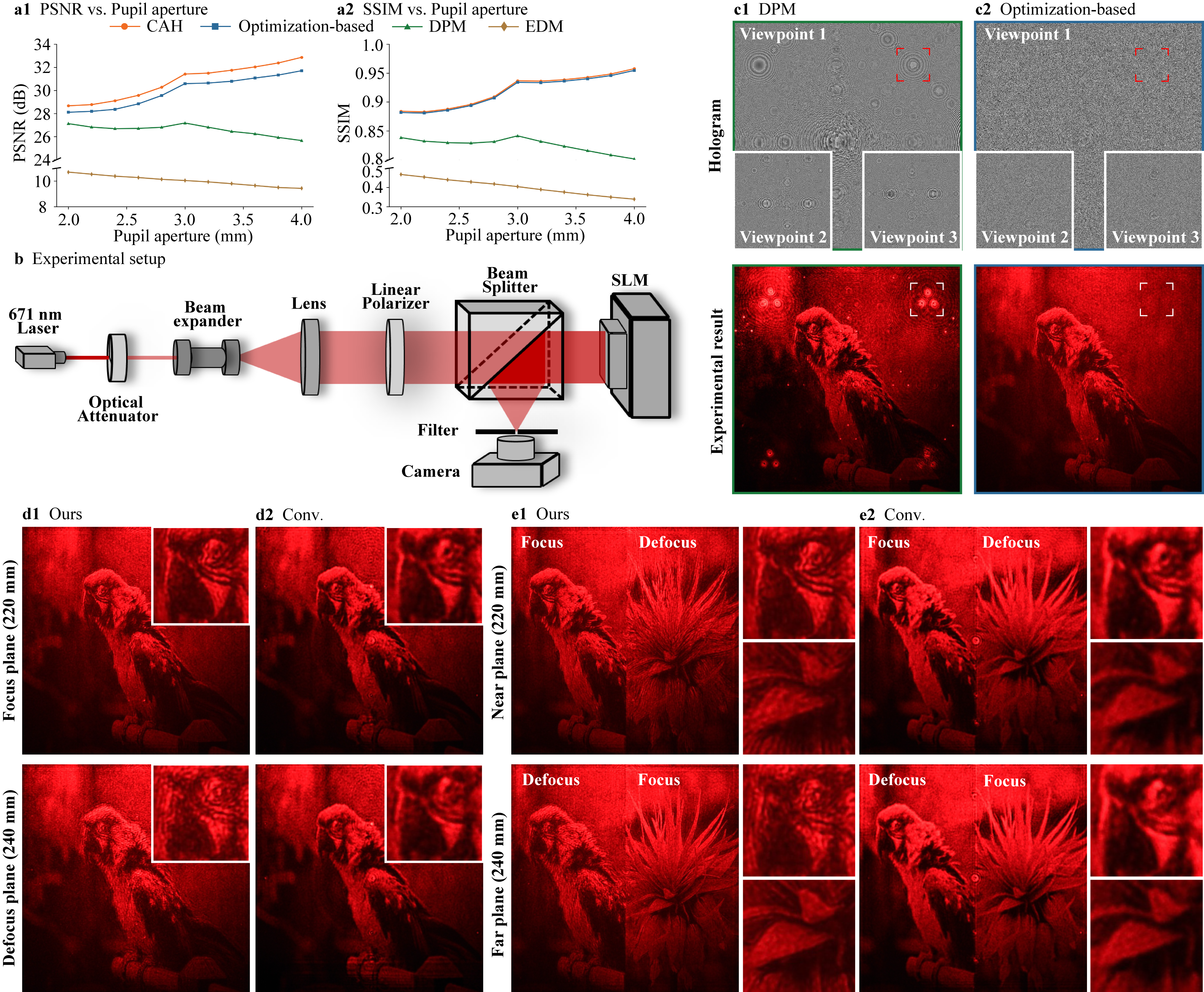}
    \caption{Comprehensive optical evaluation and benchmarking of the proposed holographic NED method. (a1,a2) Quantitative comparison of reconstruction performance across different phase encoding methods (optimization-based, DPM, and EDM) evaluated across varying pupil apertures. (b) Schematic depiction of the experimental setup. (c) POHs generated using (c1) DPM and (c2) Optimization‐based encoding technique, alongside their corresponding optical reconstructions. (d) Comparative optical reconstructions at focus plane (\SI{220}{\milli\meter}) and defocus plane (\SI{240}{\milli\meter}): (d1) Proposed method; (d2) Conventional spherical phase holography. (e) Demonstration of optical biplane image reconstruction on two distinct depth planes ("parrot" at \SI{220}{\milli\meter} and "fritillaria imperialis" at \SI{240}{\milli\meter}): (e1) Proposed method; (e2) Conventional spherical phase holography.}
    \label{fig4}
\end{figure*}

Figure \ref{fig4}(a) presents a comparative analysis of different phase encoding techniques at varying pupil apertures. 
The optimization-based encoding technique introduced herein demonstrated marked superiority over DPM and EDM, particularly evident at larger apertures.
Figure \ref{fig4}(b) depicts the experimental setup, comprising a CAS MICROSTAR FSLM-2K70-P02 spatial light modulator (SLM) with a resolution of $1920 \times 1080$ pixels and a pixel pitch of \SI{8}{\micro\meter}, illuminated by an \SI{671}{\nano\meter} all-solid-state laser. Image acquisition was performed using a Nikon D810 DSLR camera equipped with a Nikon AF-S VR 105mm f/2.8G IF-ED lens, with a \SI{2}{\milli\meter} aperture filter positioned in front of the lens to simulate the human eye’s pupil. 
A finding of this study is highlighted in Figure \ref{fig4}(c), where POHs generated using DPM exhibit significant spherical phase features, which introduce optical artifacts that detract from the overall image quality, seriously affecting the viewing experience. In contrast, the proposed optimization-based encoding technique effectively mitigates these artifacts, resulting in superior optical reconstructions. Consequently, all subsequent experimental evaluations employed this optimization-based encoding technique.

Figure \ref{fig4}(d) presents an optical comparison between the proposed and conventional methods, demonstrating that the proposed method effectively preserves high-frequency details at the focus plane while enhancing defocus blur at defocus planes. These improvements contribute to superior image quality and more pronounced depth perception compared to conventional methods.

To further validate the effectiveness of the proposed method in 3D display applications, we conducted biplane image reconstructions featuring two distinct depth layers: "parrot" at \SI{220}{\milli\meter} (near plane) and "fritillaria imperialis" at \SI{240}{\milli\meter} (far plane), as illustrated in Figure \ref{fig4}(e). The proposed method effectively delivers clear and pronounced depth cues, where refocusing on one plane induces perceptual blurring of the other, thereby reinforcing depth realism. Conversely, conventional spherical phase holography exhibits diminished depth discrimination, as both planes remain relatively sharp under focus adjustments. The superiority of the proposed approach in preserving high-frequency details and accurately simulating depth-induced blur is further substantiated through zoomed-in comparative evaluations.

\section{Discussion}\label{sec:Discussion}
\begin{figure*}[h]
    \centering
    \includegraphics[width=6.4in]{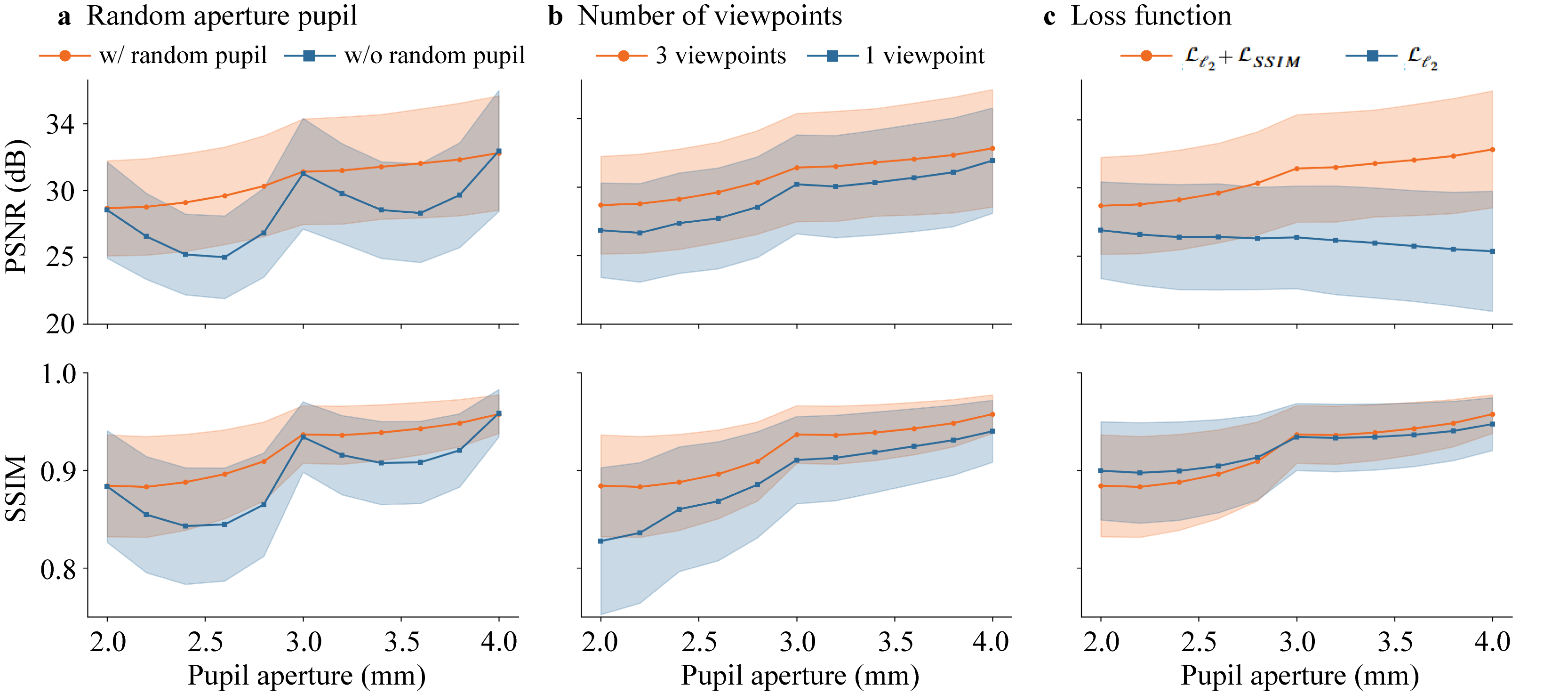}
    \caption{Quantitative analysis of factors influencing holographic reconstruction quality. (a) Effect of the random aperture pupil: with versus without random aperture pupil. (b) Effect of the number of viewpoints: three viewpoints versus one viewpoint. (c) Effect of the loss function strategy: $\ell_2$ alone versus combined $\ell_2$ and SSIM.}
    \label{fig5}
\end{figure*}

A detailed analysis was conducted to elucidate the influence of key experimental conditions on holographic reconstruction quality, as presented in Figure \ref{fig5}.

Figure \ref{fig5}(a) highlights the significance of employing the random aperture pupil on reconstruction quality. The absence of the random aperture pupil results in substantial degradation in reconstruction quality whenever the pupil aperture deviates from the predetermined fixed aperture, thereby limiting optimal image quality to specific configurations. In contrast, integrating random aperture pupils consistently preserves superior image quality across the complete physiological pupil aperture range. This demonstrates the robustness and adaptability of the proposed method, significantly mitigating adverse effects caused by physiological variations inherent within the human visual system.

Figure \ref{fig5}(b) examines the effect of the number of viewpoints on reconstruction quality. The use of three distinct viewpoints improved quality compared to a single-viewpoint configuration. This improvement can be attributed to richer complementary spatial information obtained from multiple viewpoints.
Additionally, rays originating from distinct viewpoints fail to converge accurately on the retina under defocus conditions, effectively enhancing depth perception cues. 
These findings indicate the superiority of multi-viewpoint configurations for achieving realistic and precise holographic visualization.

Figure \ref{fig5}(c) provides an evaluation of optimization strategies via different loss functions. Reconstruction employing solely an $\ell_2$ loss exhibited inferior visual fidelity compared to a combined loss function integrating both $\ell_2$ and SSIM. The incorporation of SSIM markedly improved reconstruction quality, particularly enhancing structural preservation and perceptual accuracy, thereby demonstrating the significant advantage conferred by the combined optimization strategy.

Despite the demonstrated effectiveness of our optimization-based phase encoding technique in mitigating spherical artifacts during optical reconstructions, incidental speckle noise remains an inherent challenge. Consequently, further exploration and development of advanced phase encoding techniques to achieve an eyepiece-free holographic NED free from speckle noise and artifact remains a compelling and significant research avenue.

\section{Conclusion}
In conclusion, this study introduces a novel eyepiece-free pupil-optimized holographic NED, which effectively addresses critical challenges posed by finite and dynamically varying pupil apertures in practical holographic NED systems. By employing customized spherical phase modulation to generate multiple viewpoints within the pupil, our method entirely circumvents the complexities and aberrations inherent in traditional relay optics and eyepiece-based solutions. The integrated co-optimization of amplitude and phase distributions across these viewpoints further mitigates image degradation due to finite pupil sampling, significantly enhancing visual quality and depth realism. 
Additionally, our proposed optimization-based POH encoding technique provides a high-quality, artifact-suppressed phase-only representation conducive to the practical optical implementation of eyepiece-free holographic NED. Extensive numerical validations and experimental demonstrations underscore the effectiveness of our method, consistently delivering stable, high-quality 3D imagery across diverse viewing conditions. Ultimately, this work represents a crucial advancement toward the realization of compact, lightweight, and flexible holographic NEDs, paving the way for their practical deployment in wide VR/AR applications.


\bibliography{main}

\end{document}